\documentclass[twocolumn,showpacs,preprintnumbers,amssymb]{revtex4}
\usepackage{pdfpages}
\usepackage{graphics}
\usepackage{epsfig} 
\usepackage{hyperref}
\usepackage{graphics}
\usepackage{color}      
\usepackage{graphicx}
\usepackage{graphicx}
\usepackage{dcolumn}

\begin{document}

\title{Entropy Production and Thermodynamic Dynamics in Active and Passive Brownian Systems Driven by Time-Dependent Forces and Temperatures
}
\author{Mesfin Asfaw  Taye}
\affiliation {West Los Angeles College, Science Division \\9000  Overland Ave, Culver City, CA 90230, USA}

\email{tayem@wlac.edu}

\begin{abstract}
   In this work, we  examine the  impact of time-varying temperature and force  on the thermodynamic features of active Brownian motor  that moves with velocity $v_0$ against the force as well as passive  Brownian motor.  By deriving analytical expressions for entropy production and  entropy extraction rates, we extend the existing theoretical frameworks  by considering   a force or temperature that varies  exponentially, linearly, and quadratically. By studying the  system analytically,  we investigate  how  thermal relaxation, steady-state  conditions, and nonlinear dissipation effects are affected over time. We find  that the  total entropy depends only on temperature and viscous friction if the Brownian particle moves freely, while the entropy production and dissipation rates are strongly influenced by the external force. When a Brownian particle is exposed to  periodic forcing, entropy production exhibits oscillatory behavior with monotonic decay, whereas periodic impulsive forces induce discrete spikes followed by relaxation, reflecting intermittent energy injection and dissipation. On the other hand, nonperiodic impulsive forces lead to abrupt entropy surges, followed by gradual stabilization, ensuring long-term equilibration.  At the stall force, when $f = \gamma v_0$, all thermodynamic rates—including the entropy production  and extraction rates vanishes.
We believe that our results  have    broad implications for the optimization of molecular motors, nanoscale transport, and pulse-driven systems. It also provides insights into  the design  of bio-inspired nanomachines, thermodynamically controlled microfluidic devices, and artificial nanorobots.  \end{abstract}
\pacs{Valid PACS appear here}
\maketitle


\section{Introduction}

Most active-matter systems \cite{mmg1,mmg2,mmg3,mmg4} and molecular motors operate far from equilibrium by irreversibly consuming energy and generating entropy. From intracellular transport driven by kinesin and myosin to the locomotion of bacteria and synthetic colloids, these systems achieve directed motion by violating the detailed balance. Therefore, optimizing their efficiency and velocity is crucial. The Fokker-Planck equation provides a foundational framework for capturing the stochastic dynamics  of such systems, and  the entropy production and energy dissipation   rate are governed by    drift and diffusion governing terms. In realistic environments, the  temperature, viscous friction, and external forces are time-dependent. Thus,   the variations in these parameters   reshape the energy landscape, which may lead  to intricate entropy production and extraction behaviors that defy steady-state approximations.  In this work, we systematically investigate how time-dependent  force or temperature  influences system dynamics.  This, in turn,  bridges a crucial gap by elucidating the thermodynamics of active systems under fluctuating conditions. Previous studies have explored parameter dependence in isothermal and spatially varying thermal environments \cite{mu1,mu2,mu3,mu4,mu5,mu6,mu7,mu8,mu9,mu10,mu11,mu12,ta1,mu13,mu14,mu15,mu16,mu17,muu17,mu25,mu26,mu27}, including both classical \cite{mu17,muu17} and quantum regimes \cite{mu25,mu26,mu27}, using discrete master equations \cite{mu1,mu2,mu3,mu4,mu5,mg41} and continuous Fokker-Planck formalisms \cite{mu7,mu8,ta1,muuu17,muuu177}. These studies have provided  a basic  understanding of entropy production through stochastic thermodynamics \cite{mu6}, time-reversal symmetries \cite{mar2,mar1}, fluctuation theorems \cite{mg6,mg7,mg8}, and thermodynamic uncertainty relations (TURs) \cite{mg10,mg11,mg12,mg14,mg15}. Notably, recent extensions to non-Markovian systems have revealed that memory effects can fundamentally alter non-equilibrium behavior \cite{mg9}.

How do molecular machines maintain efficiency when the forces and temperatures driving them are constantly shifting? While much of the existing research has focused on isothermal or spatially varying environments, real-world systems, from intracellular motors to synthetic nanodevices, operate under forces and temperatures that fluctuate in time.  This, in turn,  significantly affects transport efficiency, dissipation, and entropy production. In this study, we analytically investigate the thermodynamic behavior of active Brownian motor  that moves with velocity $v_0$ against the force as well as passive  Brownian motor under time-dependent thermal and force protocols. By deriving expressions for entropy, entropy production, and extraction rates for exponentially, linearly, and quadratically varying forces and temperatures, we reveal transient deviations from the equilibrium and nonlinear dissipation effects. We find that total entropy depends only on temperature  (only when the particle moves freely without boundary condition) and viscous friction, while entropy production and dissipation are force-sensitive. Periodic forces generate oscillatory entropy production with monotonic decay; periodic impulses induce discrete spikes, and nonperiodic impulses lead to abrupt entropy surges, followed by gradual stabilization. At the stall force, when $f = \gamma v_0$, all thermodynamic rates, including the entropy production  and extraction rates, vanish.

The rest of the paper is organized as follows: In Section II, we present the model and derivation of the free energy.  In Section III, considering a time-varying  temperature, we explore the dependence of the thermodynamic relations on the model  parameters.   In Section IV, we study model systems  that are subjected to time-varying forces.  Section V deals with   time-dependent    viscous friction.  In Section VI, we discuss the optimization procedure  for small-molecule motors.   Section VII  presents a summary and conclusions.

\section{The Model and  Derivation of Thermodynamic Relations}   

Before  exploring  
the effect  of time varying force and temperature,  let us derive some of the thermodynamic
 relations that are relevant to active matter, molecular motors, and artificial motors. 
The Fokker-Planck equation  \begin{equation} \frac{\partial P(x, t)}{\partial t} = -\frac{\partial}{\partial x} \left[ {A(x, t) P(x, t)\over \gamma(t)} \right] + \frac{\partial^2}{\partial x^2} \left[ {D(t) P(x, t)\over \gamma(t)} \right], \end{equation}
 governs  system dynamics. Here, $A(x, t)$ denotes the drift coefficient, which captures the effective force acting on the Brownian particle. When the particle moves autonomously with a self-propulsion velocity $v_0$, it functions as an active Brownian motor. In this case, the drift becomes $A(x,t) = f - \gamma v_0$, where $f$ represents the external load. Alternatively, when the particle undergoes unidirectional motion solely due to thermal asymmetry or an applied load without self-propulsion, it operates as a passive Brownian motor with $A(x,t) = f$. On the other hand,   $D(t)$ denotes the 
time-dependent diffusion term. The diffusion coefficient $D(t)$ is related to 
 temperature via the Einstein relation: \begin{equation} D(x, t) = \frac{k_B T(t)}{\gamma(t)}, \end{equation} where $k_B$ 
 denotes  the Boltzmann constant, and $\gamma( t)$ represents  the friction coefficient.   
  Moreover, the probability flux is given by  
	\begin{equation} J(x, t) = {A(x, t) P(x, t)\over \gamma(t)} - {D( t)\over \gamma(t)} \frac{\partial P(x, t)}{\partial x}. \end{equation}

	The entropy $S(t)$ can be   written  as  \begin{equation} S(t) = -\int P(x, t) \ln P(x, t) \, dx. 
	\end{equation}  The rate of change of entropy becomes
	\begin{equation} \frac{d S(t)}{dt} = \dot{e}_p - \dot{h}_d, \end{equation} where: 
	\begin{eqnarray} \dot{e}_p &=& \int \frac{\gamma(t) J^2}{P(x, t) T(t)} \, dx, \\ \nonumber \dot{h}_d &=& \int \left(J \frac{U'(x)}{T(t)} \right) \, dx. 
	\end{eqnarray} At steady state: \begin{equation} \frac{d S(t)}{dt} = 0  \end{equation} indicating $\dot{e}_p = \dot{h}_d > 0$.
	At the equilibrium (stationary state), the detailed balance is preserved ($J = 0$), leading to $ \dot{e}_p = \dot{h}_d = 0 $.  
	The change in  entropy (in time), entropy production, and heat dissipation are expressed as 
	\begin{eqnarray} \Delta h_d(t) &=& \int_{t_0}^{t} \dot{h}_d(t) \, dt, \\ \nonumber \Delta e_p(t) &=& \int_{t_0}^{t} \dot{e}_p(t) \, dt, \\ \nonumber \Delta S(t) &=& \Delta e_p(t) - \Delta h_d(t). \end{eqnarray}
	These relations provide a comprehensive framework for analyzing the thermodynamic behavior of systems with temporally varying  force and temperature,   highlighting the interplay between stochastic forces, thermal gradients, and entropy dynamics.

		The heat-dissipation rate $\dot{H}_d$ is related to $\dot{h}_d$. 
		After some algebra, one gets  
		\begin{eqnarray} \dot{H}_d  &=& \int \left( J U'(x)  \right) dx.
		\end{eqnarray} 
		The entropy production rate $\dot{E}_p$ is related to 
		$\dot{e}_p$ and it is given by
		\begin{equation} \dot{E}_p = \int \frac{\gamma(t)J^2}{P(x, t)} dx. 
		\end{equation} 
		The entropy balance equation for the system can be rewritten  as 
			\begin{eqnarray} \frac{dS(t)^T}{dt} &=& \dot{E}_p - \dot{H}_d. 
			\end{eqnarray} Integrating the above equation with time leads to  extensive 
			thermodynamic relations $ \Delta H_d(t) = \int_{t_0}^{t} \dot{H}_d(t) \, dt $, $\Delta E_p(t) = \int_{t_0}^{t} \dot{E}_p(t) \, dt$, and  
			$\Delta S(t)^T = \int_{t_0}^{t} \frac{dS(t)^T}{dt} \, dt$. where $ \Delta S(t)^T = \Delta E_p(t) - \Delta H_d(t) $.   
			The rate of change of internal energy  has a form  
			\begin{equation} \dot{E}_{\text{in}} = \int \left( J U'_s(x)  \right) dx \end{equation} 
			where $U'_s(x) $ denotes the internal potential. The external work rate is given by $ \dot{W} = \int J f \, dx $ where $f$ denotes the force. 
			The first law of thermodynamics  relates the above relations
				\begin{equation} \dot{E}_{\text{in}} = -\dot{H}_d - \dot{W}. 
				\end{equation} 
				
				Integrating the above equation with time, we obtain  $ \Delta E_{\text{in}} = -\int_{t_0}^{t} \left(\Delta H_d + \Delta W\right) dt. $ The free energy dissipation rate for non-isothermal systems  has a form  
				\begin{eqnarray} \dot{F} &=& \dot{E}_{\text{in}} - \dot{S}^T \\ \nonumber &=& \dot{E}_{\text{in}} - \dot{E}_p + \dot{H}_ d,
				\end{eqnarray} 
				The change in free energy  is given as 
				\begin{equation} \Delta F(t) = -\int_{t_0}^{t} \left( f V(t) + \dot{E}_p(t) \right) dt.
				\end{equation}  
			At the quasistatic limit, where $f = v_0$, both the entropy production and dissipation rates vanish; that is, $\dot{E}_p(t) = \dot{H}_d(t) = 0$, corresponding to a reversible process. Conversely, in far-from-equilibrium systems, $E_p > 0$, which reflects irreversibility. For non-isothermal systems at steady state, $ \dot{E}_p = \dot{H}_d, $ ensuring that $ \Delta F(t) = \Delta U(t). $

\section{Thermodynamic relations in   time varying  thermal arrangements}

In this section, we explore the dependence of the key  thermodynamic relations  on  the time-varying  temperature.  We  show  that the entropy production and extraction rates fluctuate over time.   In real physical systems,    these temporally varying thermal arrangements  significantly affect entropy production. For instance, rapid cooling (thermal quenching) disrupts motor efficiency, whereas gradual heating enhances efficiency but increases energy dissipation.

In this study, whenever we plot the figures,  setting  the Boltzmann constant to unity ($k_B = 1$),   we introduce the dimensionless force as $\bar{f} = \frac{f L}{T}$, where $f$ is the applied force, $L$ is a characteristic length scale (such as the motor's step size), and $T$ is the thermal energy. The rescaled temperature is given by $\bar{T} = \frac{T}{T_0}$  where $T_0$ is a reference temperature.  Before  rescaling the time, let us introduce the characteristic diffusion timescale $\tau = \frac{\gamma L^2}{T}$, and the dimensionless time is given by $\bar{t} = \frac{t T}{\gamma L^2}$. The Brownian particle is also moves autonomously  and its velocity is rescaled as $\bar{v}_0 = \frac{T}{\gamma L} v_0$. Hereafter, to avoid  confusion, the bar will be dropped.

Before deriving  the expression for entropy, let us first  rewrite the Fokker Planck equation as 
  \begin{equation} \frac{\partial P(x, t)}{\partial t} = \frac{\partial}{\partial x} \left( \frac{f'}{\gamma} P(x, t) \right) + \frac{\partial}{\partial x} \left( \frac{T(t)}{\gamma} \frac{\partial P(x, t)}{\partial x} \right) .\end{equation}  Here the motor is considered to move autonomously   with velocity $v_0$  against the load $f$  and hence $f'=f-\gamma v_0$.
	The force $f$   can be either an external force or a force generated by the active matter and the parameter  $T(t)$  denotes the time varying  temperature. 
 After some algebra,  the probability density   for  periodic boundary condition  (when the particle is walks from $x=0$ to $x=L$), is simplified to  
	\begin{equation} P(x, t) = \sum_{n=0}^\infty  \cos\left(\frac{n\pi}{L} \left(x - vt\right)\right) e^{-\mu_n(t)}, \end{equation} where $v = \frac{f-\gamma v_0}{\gamma}$  denotes  the drift velocity, $\mu_n(t) = \int_0^t \lambda_n(t') dt'$, and $\lambda_n(t) = \frac{(n\pi)^2 T(t)}{\gamma L^2}$.   In the case when the particle  diffuses  freely,   the probability density   takes a smile form 
	 \begin{equation} P(x, t) = \frac{1}{\sqrt{4 \pi \sigma^2(t)}} \exp\left(-\frac{(x - {f' t\over \gamma})^2}{4 \sigma^2(t)}\right), \end{equation} where $\sigma^2(t)$ represents the time-dependent variance.  We find that  all thermodynamic relations are independent of the boundary conditions.

\subsection{Exponentially decreasing temperature  in time} 
To explore the effect of thermal arrangements,  we introduce  exponentially and linearly decreasing thermal arrangements. 
For an exponential decay in temperature, the temperature $T(t)$ is given by
\begin{equation}
T(t) = T_{\text{st}} + (T_0 - T_{\text{st}}) e^{-\beta t},
\end{equation}
where $T_0$ is the initial temperature and $T_{\text{st}}$ is the steady-state temperature. The variance is expressed as
$
\sigma^2(t) = \frac{2}{\gamma} \left[ T_{\text{st}} t + \frac{(T_0 - T_{\text{st}})}{\beta} \left( 1 - e^{-\beta t} \right) \right].
$
For exponentially decreasing case,  the entropy is given as
 \begin{equation}
S(t) = \frac{1}{2} \ln \left( 4\pi e \frac{2}{\gamma} \left[ T_{\text{st}} t + \frac{(T_0 - T_{\text{st}})}{\beta} (1 - e^{-\beta t}) \right] \right),
\end{equation}
{\it  Please note that the expression for entropy is independent of the load $f'$ if only free boundary conditions are imposed.
 }
The change in   entropy  is given by  
\begin{eqnarray}
\Delta S(t)& =& S(t) - S(t_0) \\ \nonumber
&=& \frac{1}{2} \ln \left( \frac{T_{\text{st}} t + \frac{T_0 - T_{\text{st}}}{\beta} (1 - e^{-\beta t})}{T_{\text{st}} t_0 + \frac{T_0 - T_{\text{st}}}{\beta} (1 - e^{-\beta t_0})} \right).
\end{eqnarray}
The above equation indicates that the entropy increases with time.  
\begin{figure}[ht]
\centering
{
    \includegraphics[width=6cm]{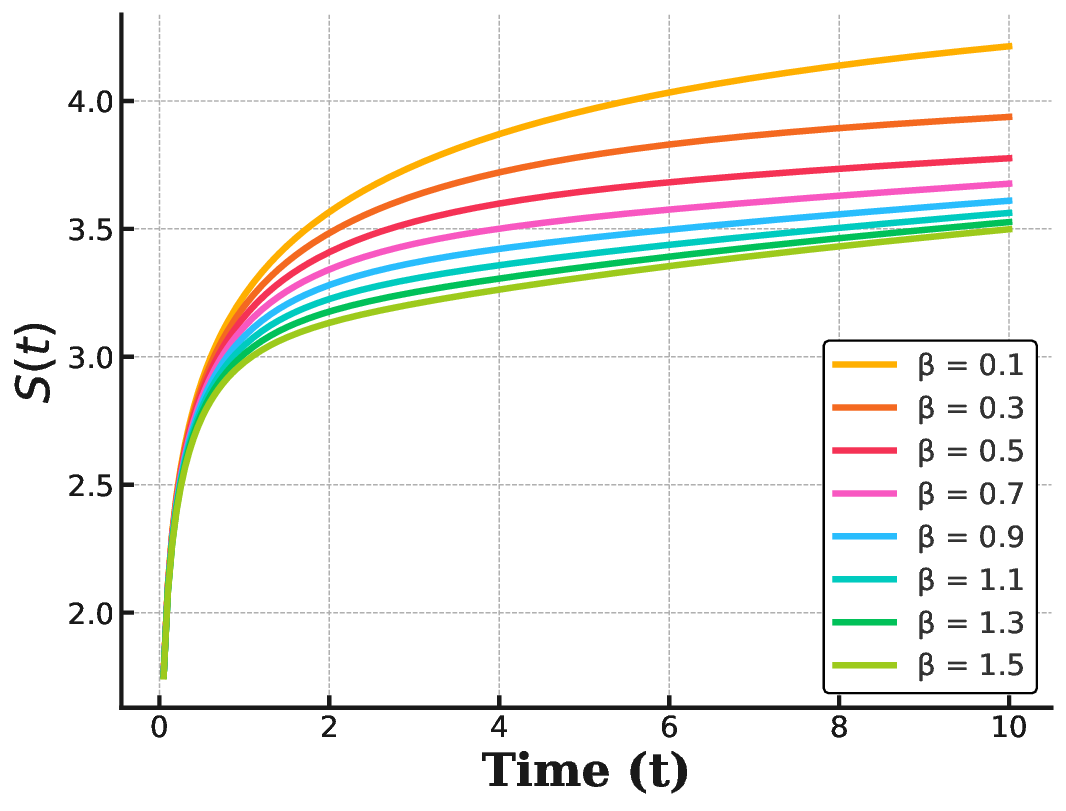}
}
\caption{ (Color online)      Entropy evolution $S(t)$  for exponential decay model   as a function of time for different values of $\beta$ ($\beta = 0.1, 0.3, 0.5, 0.7, 0.9, 1.1, 1.3, 1.5$), with $T_0 = 10$ (initial temperature) and $T_{\text{st}} = 1$ (steady-state temperature).  $v_0$    } 
\label{fig:sub} 
\end{figure}
The entropy evolution, $S(t)$  , is analyzed for the exponential decreasing temperature case for various values of the decay parameter $\beta$ ($\beta = 0.1, 0.3, 0.5, 0.7, 0.9, 1.1, 1.3, 1.5$), while maintaining the same initial and steady-state temperatures of $T_0 = 10.0$ and $T_{\text{st}} = 1.0$  (see Fig. 1).

\subsection{Linearly  decreasing temperature  in time} 
On the contrary,   for linearly decreasing  thermal arrangements
\begin{equation}
T(t) = T_0 - \alpha t
\end{equation}
which is valid   for the time interval $ t \leq \frac{T_0 - T_{\text{st}}}{\alpha} $.  We calculate the  variance  as
$
\sigma^2(t) = \frac{2}{\gamma} \left[ T_0 t - \frac{\alpha t^2}{2} \right]$.
We then calculate the entropy 
\begin{equation}
S(t) = \frac{1}{2} \ln \left( 4\pi e \frac{2}{\gamma} (T_0 t - \frac{\alpha t^2}{2}) \right),
\end{equation}
As one can see that the entropy depends only  on the temperature and viscous friction.
The steady-state entropy  can be written as 
\begin{equation}
S = \frac{1}{2} \ln \left( 4\pi e \frac{(T_0 - T_{\text{st}})^2}{\gamma \alpha} \right).
\end{equation}
The change in entropy is given by 
\begin{equation}
\Delta S(t) = \frac{1}{2} \ln \left( \frac{T_0 t - \frac{\alpha t^2}{2}}{T_0 t_0 - \frac{\alpha t_0^2}{2}} \right),  t \leq \frac{T_0 - T_{\text{st}}}{\alpha}.
\end{equation}
The entropy evolution, $S(t)$, is analyzed for the linear temperature decay model (see Fig. 2), $S(t)$ is examined for different values of the parameter $\alpha$ ($\alpha = 0.1, 0.3, 0.5, 0.7, 0.9, 1.1, 1.3, 1.5$), with an initial temperature of $T_0 = 10.0$ and a steady-state temperature of $T_{\text{st}} = 1$.  
\begin{figure}[ht]
\centering
{
    \includegraphics[width=6cm]{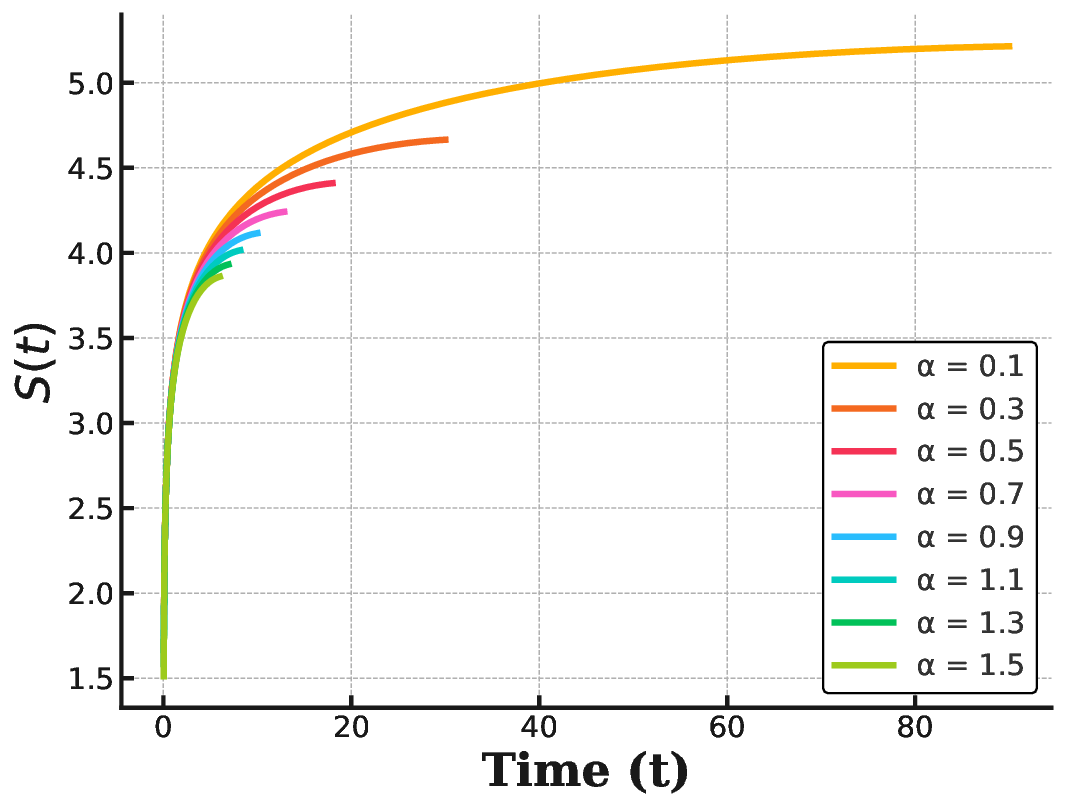}}
\caption{ (Color online) Entropy evolution $S(t)$ as a function of time ( linear temperature decay model)  for different values of $\alpha$ ($\alpha = 0.1, 0.3, 0.5, 0.7, 0.9, 1.1, 1.3, 1.5$), with $T_0 = 10$ (initial temperature) and $T_{\text{st}} = 1$ (steady-state temperature).  } 
\label{fig:sub} 
\end{figure}

Regardless of any thermal arrangements,   the current density  can be evaluated via 
\begin{equation}
J(x, t) = P(x, t) v, \quad v = \frac{f-\gamma v_0}{\gamma},
\end{equation}
After some algebra, for all thermal profiles, we derive the expressions for entropy production rate 
\begin{equation}
{\dot e}_p = \frac{(f-\gamma v_0)^2}{\gamma T(t)} + \frac{T(t)}{2 \sigma^2(t)}
\end{equation}
and extraction rate 
\begin{equation}
\dot{h}_{p} = \frac{(f-\gamma v_0)^2}{\gamma T(t)}.
\end{equation}
\begin{figure}[ht]
\centering
{
    \includegraphics[width=6cm]{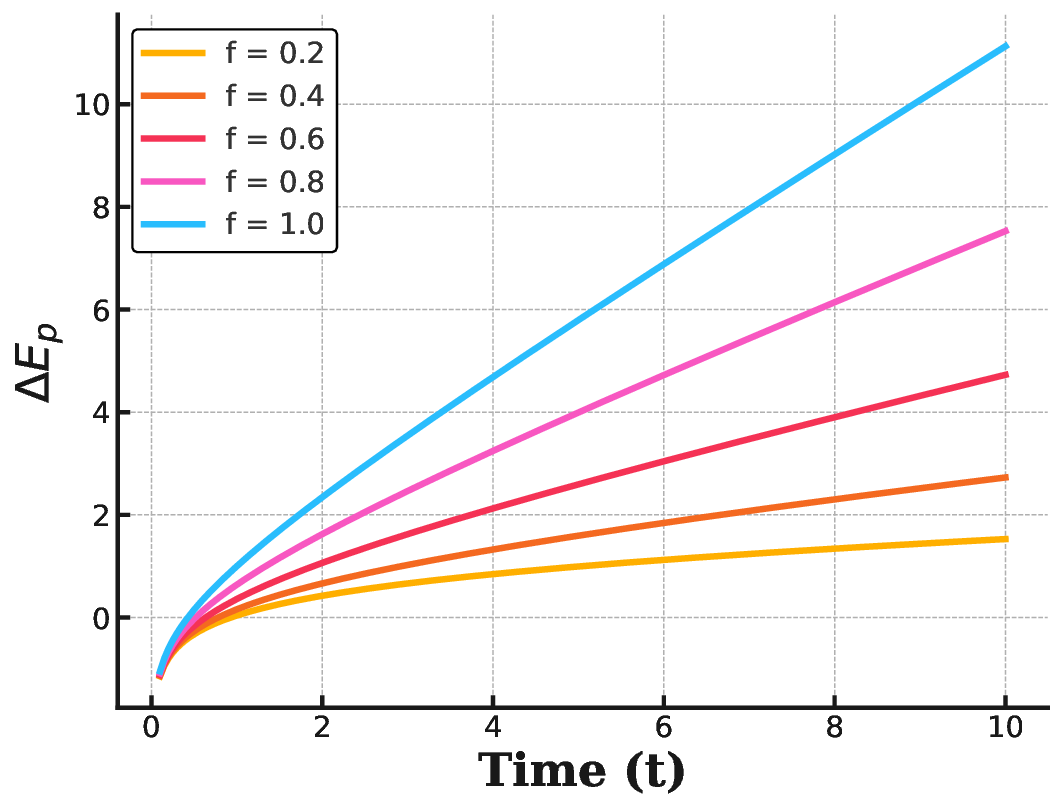}}
\caption{(Color online) The figure presents the temporal evolution of $\Delta E_p$ (in the absence of $v_0$)for various values of the force parameter $f$ ($f = 0.2, 0.4, 0.6, 0.8, 1.0$) under a linearly decreasing temperature regime, with an initial temperature of $T_0 = 10$, a steady-state temperature of $T_{\text{st}} = 1$, and a decay parameter $\alpha = 0.1$.  The result of this work also depicts that 
 $f$ enhances energy dissipation,   showing  the significant role of external forcing in the thermodynamic behavior of the system. }

\label{fig:sub} 
\end{figure} 
\begin{figure}[ht]
\centering
{
    \includegraphics[width=6cm]{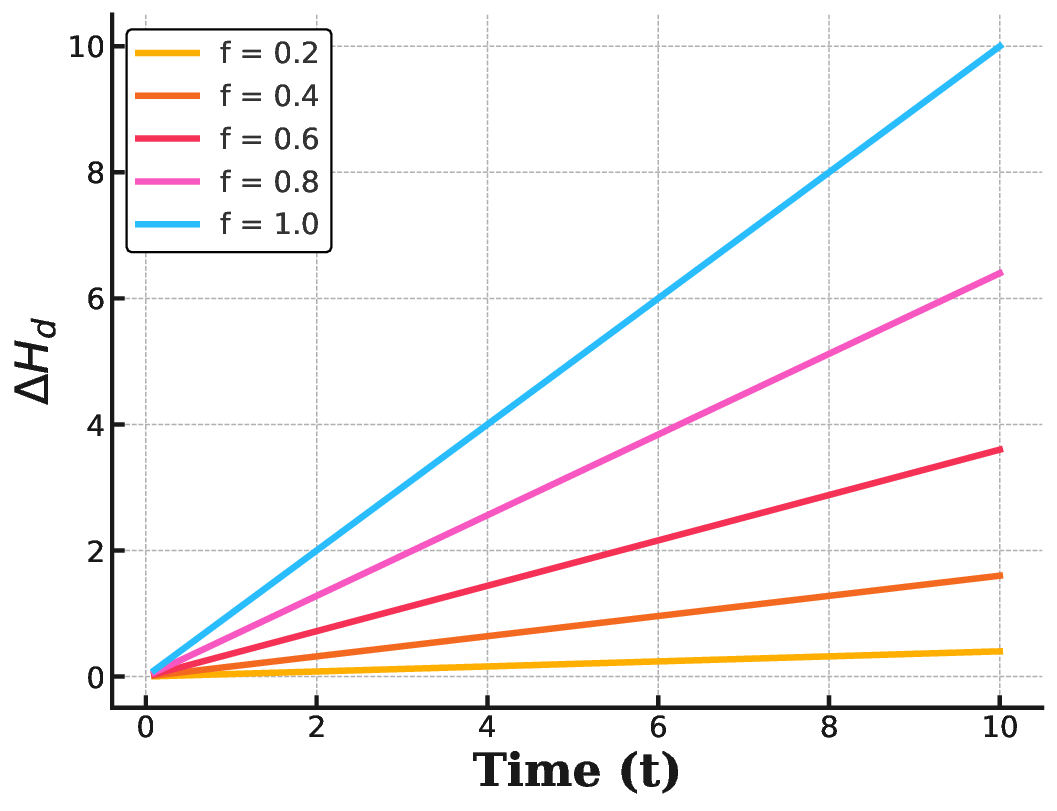}
}
\caption{(Color online)  The figure illustrates the variation of $\Delta H_d$ as a function of time (in the absence of $v_0$) for the same set of $f$ values under identical thermal conditions. The change in the heat dissipation depict that  as time increases,   this  thermodynamic   relations   increases linearly.}

\label{fig:sub} 
\end{figure} 
Fig. 3 illustrates the behavior of $\Delta E_p$ as a function of time for different values of $f$ ($f = 0.2, 0.4, 0.6, 0.8, 1.0$) under a linearly decreasing temperature profile. Similarly, Fig. 4 presents the evolution of $\Delta H_d$ over time for various values of $f$. Both figures demonstrate that the entropy production or extraction rates increase monotonically with both force and time. In the absence of an applied force, the entropy production remains zero.

Let us now derive some of the key thermodynamic quantities for the linearly decreasing cases.  The term related to  the heat extraction rate can be writen as 
\begin{equation}
\dot{H}_d = \frac{(f-\gamma v_0)^2}{\gamma}.
\end{equation}
The term related to  the entropy production rate is given by:
\begin{equation}
\dot{E}_p = \frac{(f-\gamma v_0)^2}{\gamma} + \frac{(T_0 - \alpha t)^2}{\gamma \left( 4 \left( T_0 t - \frac{\alpha t^2}{2} \right) \right)}.
\end{equation}

The time  changes in entropy production and  heat dissipation over time are
\begin{equation}
\Delta H_d = \frac{(f-\gamma v_0)^2 t}{\gamma}
\end{equation}
and 
\begin{equation}
\Delta E_p = \frac{(f-\gamma v_0)^2}{\gamma} t + \frac{1}{2} \ln \left( \frac{T_0 t - \frac{\alpha t^2}{2}}{T_0 t_0 - \frac{\alpha t_0^2}{2}} \right),
\end{equation}
which is in agreement   with $\Delta S= \Delta E_p - \Delta H_d$.  
The free energy 
\begin{equation}
\Delta F(t) = -\Delta S(t)
\end{equation}
as expected.  {\it Clearly, in the quasistatic limit, when the external force approaches the stall condition $f \to \gamma v_0$, the net velocity of the particle becomes zero. In this regime, all thermodynamic rates—including entropy production and extraction rates vanish, as shown in the equations above.}

Here, we want to emphasize that entropy plays a crucial role in understanding these systems, as it quantifies energy dissipation, irreversibility, and the approach to thermal equilibrium. Brownian ratchets and stochastic engines, which rely on thermal noise to generate directed motion, exhibit reduced performance as the temperature drops, limiting the available thermal fluctuations necessary for movement. Active matter systems, such as bacterial colonies, artificial microswimmers, and cytoskeletal assemblies, rely on energy input to sustain motion; however, under decreasing temperatures, their activity diminishes, affecting their ability to self-organize and perform functions. 
By studying how systems behave under different temperature decay patterns, researchers can develop energy-efficient molecular machines, refine biomedical technologies, enhance industrial cooling strategies, and optimize quantum and nanoscale devices.

\section{Analysis of Brownian Motion Under a Force that Depends on Time}

In this section, we show that time-varying forces play a significant role in nonequilibrium thermodynamics by dictating transport efficiency, energy dissipation, and entropy production. We show that when the magnitude of these forces increases, the entropy production and extraction rates increase. We want to stress that most biological systems such as molecular motors, Brownian motors, active matter, and self-propelled particles either generate time-varying forces or operate under time-varying forces.

\subsection{Exponentially Decreasing Force in Time}

Let us now consider a force that exponentially decays in time and approaches a constant value at steady state:
\begin{equation}
F(t) = f_{st} - \gamma v_0 + (F_0 - f_{st}) e^{-\lambda t}
\end{equation}
where $F_0$ denotes the initial force amplitude, $f_{st}$ denotes the steady-state force before adjusting for stall, $\gamma v_0$ is the stall force, and $\lambda$ represents the decay rate.

Considering constant temperature and the adjusted exponentially decreasing force, we rewrite the Fokker-Planck equation as:
\begin{eqnarray}
\frac{\partial P(x,t)}{\partial t} &=& \frac{\partial}{\partial x} \left( \frac{f_{st} - \gamma v_0 + (F_0 - f_{st}) e^{-\lambda t}}{\gamma} P(x,t) \right)\\ \nonumber && + \frac{T}{\gamma} \frac{\partial^2 P(x,t)}{\partial x^2}.
\end{eqnarray}
Note that this force can either be an external force applied to the system or a self-generated force arising from active matter.

After some algebra, we also find the probability distribution:
\begin{equation}
P(x,t) = \frac{1}{\sqrt{4\pi D t}} \exp \left( -\frac{\left( x - \frac{F_0}{\gamma \lambda} (1 - e^{-\lambda t}) - \frac{(f_{st} - \gamma v_0) t}{\gamma} \right)^2}{4 D t} \right).
\end{equation}
Using the above probability distribution, the entropy for the given system is simplified to:
\begin{equation}
S(t) = \frac{1}{2} \ln \left( \frac{4\pi e T t}{\gamma L^2} \right)
\end{equation}
Exploiting Eq. (37), one can see that as time progresses, the entropy increases, showing the system irreversibility increases with time. Since the external energy of the system is zero, via Eq. (14), it is evident that:
\begin{equation}
\Delta F(t) = -\Delta S(t).
\end{equation}
The change in the free energy decreases with time.

The entropy production rate is given by:
\begin{equation}
\dot{e}_p = \frac{1}{2 T t} + \frac{(F_0 - f_{st})^2}{\gamma T} e^{-2\lambda t} + \frac{(f_{st} - \gamma v_0)^2}{\gamma T}
\end{equation}
At steady state ($t \to \infty$), the entropy production rate becomes:
\begin{equation}
\dot{e}_p = \frac{(f_{st} - \gamma v_0)^2}{\gamma T}
\end{equation}
These equations depict that at small time $t$ the entropy production rate is considerably high and as time progresses, it decreases and stabilizes at a steady value dictated by the adjusted steady-state force. As long as a nonzero net force is imposed or operates in finite time, the system continuously dissipates energy.

The entropy extraction rate is calculated as:
\begin{equation}
J(x, t) = -\frac{T}{\gamma} \frac{\partial P}{\partial x} + \frac{f_{st} - \gamma v_0 + (F_0 - f_{st}) e^{-\lambda t}}{\gamma} P
\end{equation}
\begin{equation}
\dot{h}_d = \int \left( J \frac{F(t)}{T} \right) dx
\end{equation}
where the probability current is given above.

After some algebra, one gets:
\begin{equation}
\dot{h}_d = \frac{(F_0 - f_{st})^2}{\gamma T} e^{-2\lambda t} + \frac{(f_{st} - \gamma v_0)^2}{\gamma T}
\end{equation}
At steady state ($t \to \infty$), the entropy production and extraction rates converge to:
\begin{equation}
\dot{e}_p = \dot{h}_d = \frac{(f_{st} - \gamma v_0)^2}{\gamma T}
\end{equation}
This result indicates that at long times, the system reaches a balance where the entropy production rate matches the entropy extraction rate.

The velocity of the particle moving in this force field is given as:
\begin{equation}
v = \frac{f_{st} - \gamma v_0 + (F_0 - f_{st}) e^{-\lambda t}}{\gamma}
\end{equation}
and at steady state one gets $v = \frac{f_{st} - \gamma v_0}{\gamma}$.

All of these results indicate that while the entropy $S(t)$ remains independent of the applied force, the entropy production and extraction rates significantly depend on the shifted force $f - \gamma v_0$. As the force increases beyond the stall point, the system exhibits greater irreversibility, reflecting higher dissipation and entropy generation. At the stall force, when $f = \gamma v_0$, or equivalently when $f - \gamma v_0 = 0$, all thermodynamic rates  including entropy production and extraction vanish.

\subsection{Brownian Motion in a Periodic Force}

Let us next consider a force that varies periodically:
\begin{equation}
F(t) = F_0 \cos(\omega t + \phi) - \gamma v_0
\end{equation}
Here $F_0$, $\omega$, and $\phi$ denote the force amplitude, the angular frequency, and the phase shift, respectively.

We write the corresponding Fokker-Planck equation as:
\begin{equation}
\frac{\partial P(x,t)}{\partial t} = \frac{\partial}{\partial x} \left( \frac{F_0 \cos(\omega t + \phi) - \gamma v_0}{\gamma} P(x,t) \right) + \frac{T}{\gamma} \frac{\partial^2 P(x,t)}{\partial x^2}
\end{equation}
and one can simplify this equation to:
\begin{equation}
\frac{\partial P}{\partial t} = -\frac{F_0 \cos(\omega t + \phi) - \gamma v_0}{\gamma} \frac{\partial P}{\partial x} + \frac{T}{\gamma} \frac{\partial^2 P}{\partial x^2}
\end{equation}
The probability distribution for a given system is given by:
\begin{equation}
P(x,t) = \frac{1}{\sqrt{4\pi D t}} \exp \left( -\frac{\left( x - \frac{F_0}{\gamma \omega} \sin(\omega t + \phi) - v_0 t \right)^2}{4 D t} \right)
\end{equation}

After some algebra, one finds the entropy as:
\begin{equation}
S(t) = \frac{1}{2} \ln \left( \frac{4\pi e T t}{\gamma L^2} \right)
\end{equation}
which is independent of force.

The entropy production rate simplifies to:
\begin{equation}
\dot{e}_p = \frac{1}{2 T t} + \frac{(F_0 \cos(\omega t + \phi) - \gamma v_0)^2}{\gamma T}
\end{equation}
The first term, $\frac{1}{2Tt}$, represents a dissipative contribution that decays over time, while the second term, $\frac{(F_0 \cos(\omega t + \phi) - \gamma v_0)^2}{\gamma T}$, accounts for entropy fluctuations due to periodic driving. Although entropy fluctuates periodically, it decreases over time.

The time-averaged entropy production rate is also given by:
\begin{equation}
\langle \dot{e}_p \rangle = \frac{1}{2\pi} \int_0^{2\pi} \frac{(F_0 \cos(\theta) - \gamma v_0)^2}{\gamma T} d\theta = \frac{F_0^2 + \gamma^2 v_0^2}{2\gamma T}
\end{equation}

\begin{figure}[ht]
\centering
\includegraphics[width=6cm]{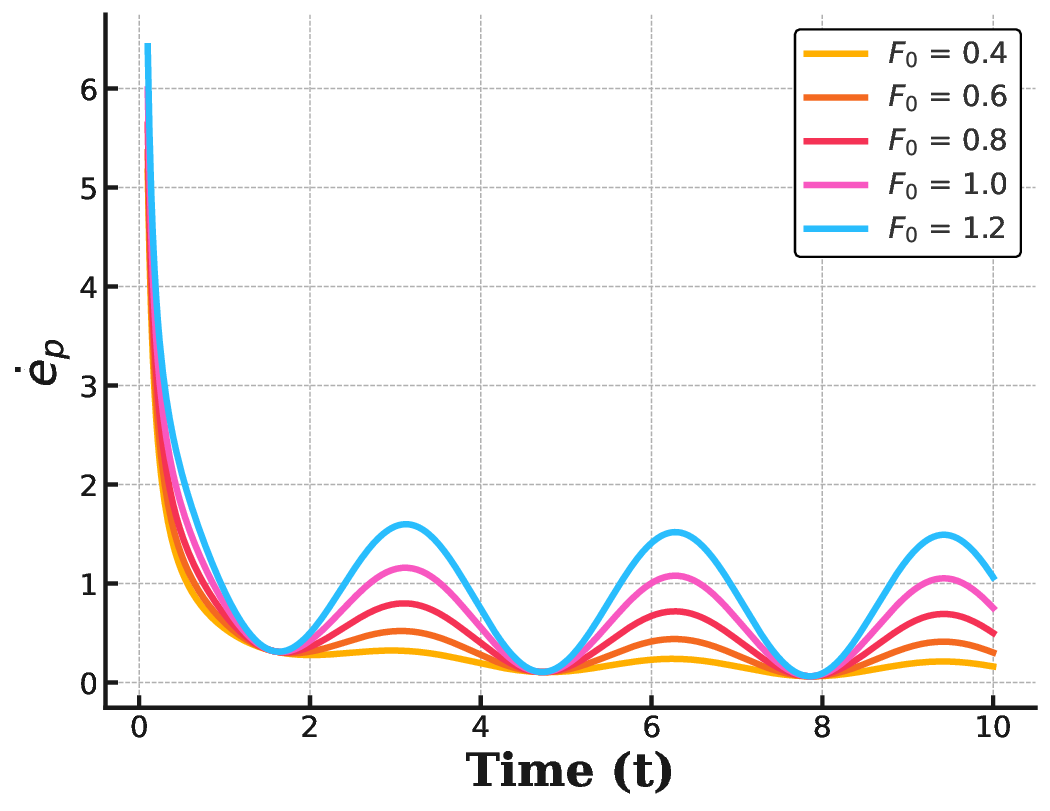}
\caption{(Color online) The figure depicts the temporal evolution of the entropy production rate, $\dot{e}_p$, for different values of the force parameter $F_0$ ($F_0 = 0.4, 0.6, 0.8, 1.0, 1.2$), with fixed parameters  $v_0=0$, $\omega = 10.0$, $T = 1.0$, $\gamma = 1$, and $\varphi = 0$. The results show that $\dot{e}_p$ exhibits an oscillatory behavior while decreasing over time. Notably, larger values of $F_0$ lead to a considerable increase in the amplitude of oscillations, indicating a stronger influence of the applied force on entropy dynamics. Despite these fluctuations, the entropy production rate declines on average at a nearly constant rate, underscoring the interplay between external forcing and dissipative thermodynamic processes.}
\label{fig:sub}
\end{figure}

The heat dissipation rate is calculated as:
\begin{equation}
\dot{h}_d = \frac{(F_0 \cos(\omega t + \phi) - \gamma v_0)^2}{\gamma T}
\end{equation}

At steady state, the time-averaged entropy production rate and heat dissipation rate are equal:
\begin{equation}
\langle \dot{e}_p \rangle = \langle \dot{h}_d \rangle = \frac{F_0^2 + \gamma^2 v_0^2}{2 \gamma T}
\end{equation}
This result indicates that at steady state, the system reaches a balance where the entropy production rate balances the entropy extraction rate.

The velocity:
\begin{equation}
v = \frac{F_0 \cos(\omega t + \phi) - \gamma v_0}{\gamma}
\end{equation}
oscillates with the same frequency $\omega$, and the time-averaged velocity is zero. For instance, if a charge is exposed to a time-varying electric field, the particle experiences a time-varying force. As a result, the particle will have oscillatory velocity.

In Fig. 5, we present the time evolution of the entropy production rate, $\dot{e}_p$ (in the absence of $v_0$), for different values of the external force amplitude $F_0$ ($F_0 = 0.4, 0.6, 0.8, 1.0, 1.2$). The results reveal that the entropy production rate exhibits an oscillatory behavior while simultaneously decreasing over time. Notably, for larger values of $F_0$, the amplitude of these oscillations becomes significantly more pronounced, indicating a stronger influence of the applied force on the system’s thermodynamic response. Despite these oscillations, the entropy production rate exhibits an overall monotonic decline, with an average decay that follows a nearly constant rate.

As discussed before, periodic force may play a crucial role in active matter, molecular motors, and thermodynamic heat engines. For instance, molecular motors get periodic energy input from ATP hydrolysis to generate mechanical work as well as to overcome thermal fluctuations and molecular friction.

\subsection{Analysis of Brownian Motion Under a Linearly Increasing Force}

In this section, we consider a linearly increasing force which can be found in real systems such as molecular motors, Brownian ratchets, and active matter. Even in active matter, self-propelled particles experience gradually increasing forces due to controlled chemical gradients or time-dependent propulsion. In Brownian ratchets, we can apply external fields, such as optical or magnetic forces, that increase over time to drive directed motion.
This force
\begin{equation}
F(t) = k t - \gamma v_0
\end{equation}
can also be used as a probe to study a given system. Here, $k$ is a constant.

Using the Fokker-Planck equation:
\begin{equation}
\frac{\partial P(x,t)}{\partial t} = \frac{\partial}{\partial x} \left( \frac{k t - \gamma v_0}{\gamma} P(x,t) \right) + \frac{T}{\gamma} \frac{\partial^2 P(x,t)}{\partial x^2},
\end{equation}
one can solve the probability distribution:
\begin{equation}
P(x,t) = \frac{1}{\sqrt{4\pi D t}} \exp \left( -\frac{\left( x - \frac{k}{2 \gamma} t^2 + v_0 t \right)^2}{4 D t} \right),
\end{equation}

The entropy is given as:
\begin{equation}
S(t) = \frac{1}{2} \ln \left( \frac{4\pi e T t}{\gamma L^2} \right)
\end{equation}
As one can see, the entropy logarithmically increases with time and temperature.

For the linearly increasing force case, the velocity:
\begin{equation}
v = \frac{k t - \gamma v_0}{\gamma}
\end{equation}
linearly increases in time.

After some algebra, the entropy production rate is given by:
\begin{equation}
\dot{e}_p = \frac{1}{2 T t} + \frac{(k t - \gamma v_0)^2}{\gamma T} + \frac{k^2}{\gamma T}
\end{equation}
The entropy production rate comprises three distinct contributions: a transient decay term, time-dependent growth term, and steady-state component. The initial decay indicates that entropy production is initially high, but diminishes as the system transitions from an out-of-equilibrium state to a driven steady state. The quadratic growth term in time reflects the increasing contribution of periodic forcing, signifying the cumulative effects of sustained energy injection into the system.

On the other hand, the heat dissipation rate is calculated as:
\begin{equation}
\dot{h}_d = \frac{(k t - \gamma v_0)^2}{\gamma T} + \frac{k^2}{\gamma T}
\end{equation}

\subsection{Brownian Motion  moving in a Periodic Impulse Force}

Let us now consider a Brownian particle that walks in a periodic impulsive force characterized by sudden bursts applied at regular intervals. The system is driven out of equilibrium when such a force is applied, and this in turn creates distinct entropy production patterns influenced by impulse frequency, magnitude, and duration. This force considerably affects the dynamics of the system because, unlike continuous force, the impulse causes rapid changes in energy and dynamics. In stochastic thermodynamics and nonlinear dynamics, this force
\begin{eqnarray}
F(t) = F_0 \sum_{n=0}^{\infty} \delta(t - nT') - \gamma v_0 \sum_{n=0}^{\infty} \delta(t - nT')
\end{eqnarray}
regulates motion, enhances energy transfer, and sustains non-equilibrium steady states, making them essential for understanding driven systems under discrete perturbations. Here, the magnitude of the impulse force $F_0$ and the period of the applied force $T'$ dictate the dynamics. The delta function $\delta(t - nT')$ represents the instantaneous impulse at discrete times, $t = nT'$.

After writing the Fokker-Planck equation:
\begin{eqnarray}
\frac{\partial P(x,t)}{\partial t} &=& \frac{\partial}{\partial x} \left( \frac{(F_0 - \gamma v_0) \sum_{n=0}^{\infty} \delta(t - nT')}{\gamma} P(x,t) \right) \nonumber \\ &&+ \frac{T}{\gamma} \frac{\partial^2 P(x,t)}{\partial x^2},
\end{eqnarray}

we calculate the probability distribution as:
\begin{equation}
P(x,t) = \frac{1}{\sqrt{4\pi D t}} \exp \left( -\frac{(x - n v' T')^2}{4 D t} \right),
\end{equation}
where $v' = \frac{F_0 - \gamma v_0}{\gamma}$ and $D = \frac{T}{\gamma}$ denote the effective velocity and diffusion coefficient, respectively.

The entropy:
\begin{equation}
S(t) = \frac{1}{2} \ln \left( \frac{4\pi e T t}{\gamma L^2} \right)
\end{equation}
increases with time and temperature.

On the other hand, the velocity:
\begin{equation}
v = \frac{(F_0 - \gamma v_0) \sum_{n=0}^{\infty} \delta(t - nT')}{\gamma}
\end{equation}
will have sudden jumps at each impulse followed by exponential decay, as shown in Fig. 6a. The time average of the velocity is given as:
\begin{eqnarray}
\langle v \rangle &=& \frac{1}{T'} \int_0^{T'} v(t) , dt \\ \nonumber
&=& \frac{F_0 - \gamma v_0}{\gamma^2 T'} (1 - e^{-\gamma T'}).
\end{eqnarray}

The entropy production rate is given as:
\begin{equation}
\dot{e}_p = \frac{1}{2 T t} + \frac{(F_0 - \gamma v_0)^2}{\gamma T} \sum_{n=0}^{\infty} \delta(t - nT').
\end{equation}
It can be seen that the entropy production rate is dictated by the periodic impulsive force. Consequently, it exhibits discrete spikes at regular intervals whenever an impulsive force is applied, and after the spikes, the rate relaxes to lower values. This oscillatory behavior originates from intermittent energy injection, which temporarily increases the entropy production before the system dissipates excess energy. The instantaneous nature of external forcing is amplified by the presence of Dirac delta functions in both entropy production and heat extraction. This indicates that energy input occurs in short bursts rather than continuously. We show that even though the entropy production rate fluctuates over time, as time increases, it declines to a steady state. Regardless of the magnitude of the force, the rate decays over time.

The entropy extraction rate is also calculated as:
\begin{equation}
\dot{h}_d = \frac{(F_0 - \gamma v_0)^2}{\gamma T} \sum_{n=0}^{\infty} \delta(t - nT').
\end{equation}
At steady state, the entropy production and entropy extraction rates are equal:
\begin{equation}
\dot{e}_p = \dot{h}_d = \frac{(F_0 - \gamma v_0)^2}{\gamma T}.
\end{equation}
This result indicates that, at long times, the system reaches a balance where the entropy produced matches the heat dissipated.

\begin{figure}[ht]
\centering
\includegraphics[width=6cm]{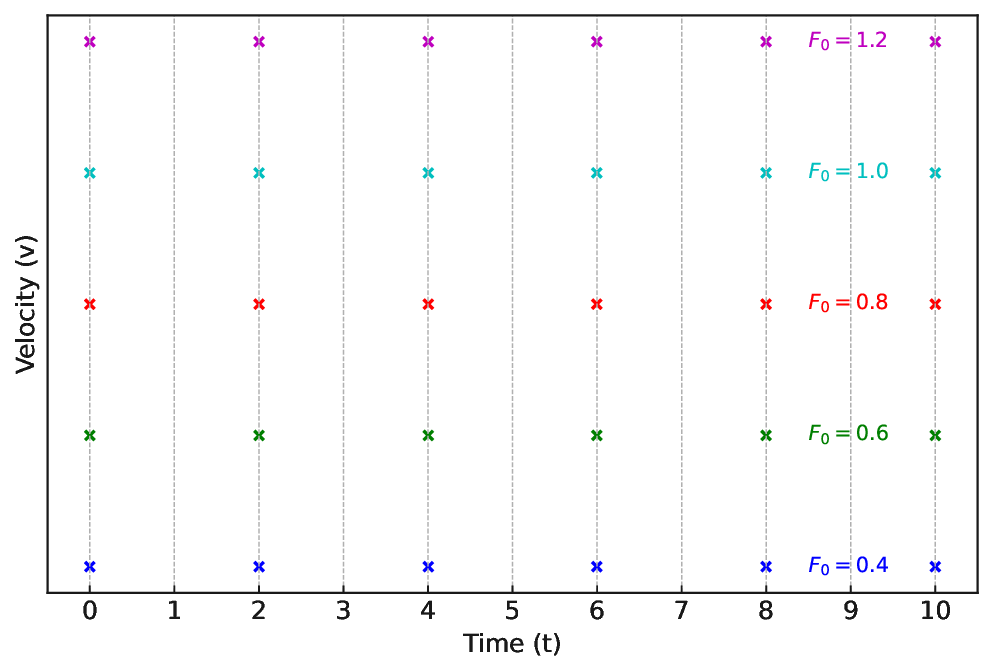}
\includegraphics[width=6cm]{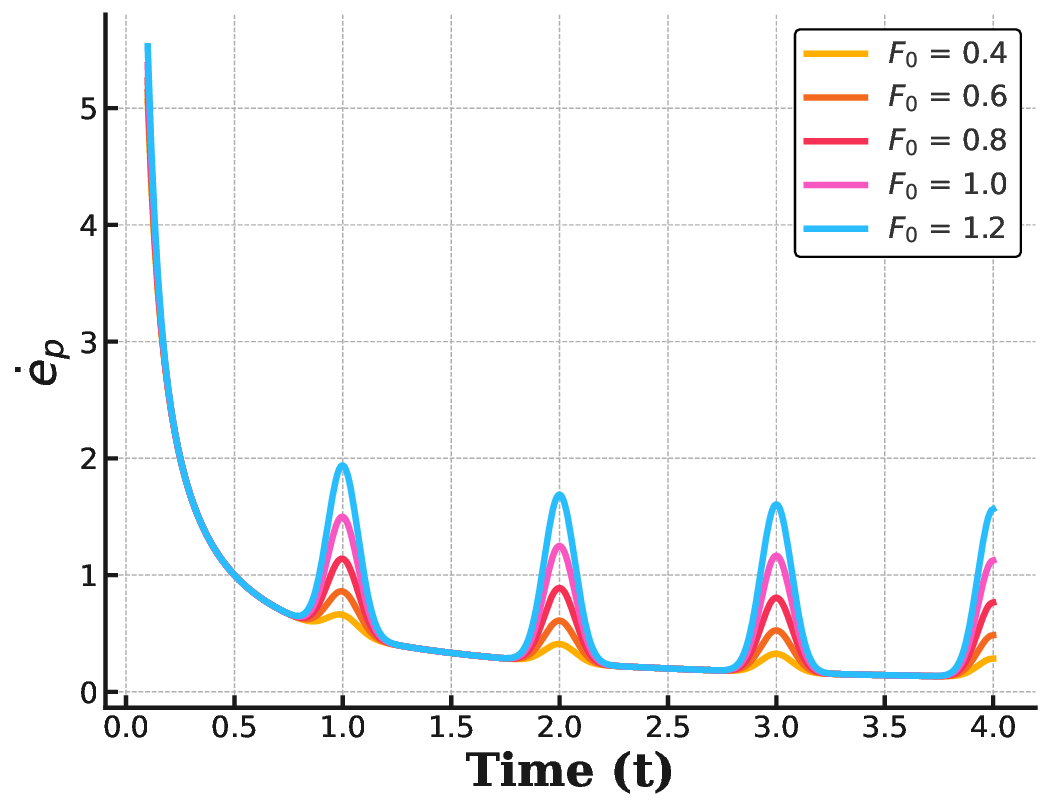}
\caption{(a) Plot of velocity as a function of time in the absence of $v_0$. (b) The plot of $\dot{e}_p$ as a function of time $t$ for different forces $F_0$ ($F_0 = 0.4, 0.6, 0.8, 1.0, 1.2$)  and $v_0=0$. The results show that whenever a periodic impulse force is applied, the entropy production rate exhibits recurrent spikes, followed by relaxation. As the magnitude of $F_0$ increases, the amplitude of these oscillations becomes more pronounced. Despite these fluctuations, the entropy production rate decreases to its steady-state value, with an average decay that remains consistent regardless of the applied force.}
\end{figure}

Figure 6a depicts $\dot{e}_p$ versus time for fixed $F_0$ ($F_0 = 0.4, 0.6, 0.8, 1.0, 1.2$)  and $v_0=0$. The figure shows that in the presence of a periodic impulse force, the entropy production rate exhibits recurrent spikes. Each spike ultimately relaxes to a lower value. This periodic modulation arises due to the external forcing, which intermittently perturbs the system’s entropy dynamics. Despite these oscillatory fluctuations, the overall entropy production rate decreases over time, following an average decay pattern.

\subsection{Brownian Motion  subjected to  Impulsive Force}

Let us now explore the thermodynamic relations in the presence of impulsive force
\begin{equation}
F(t) = F_0 \sum_{n=0}^{\infty} \delta(t - t_0) - \gamma v_0 \sum_{n=0}^{\infty} \delta(t - t_0).
\end{equation}
Fokker-Planck equation in terms of the applied force $F$ at time $t_0$ is calculated as
\begin{equation}
\frac{\partial P(x,t)}{\partial t} = -\frac{F}{\gamma} \delta(t - t_0) \frac{\partial P}{\partial x} + \frac{T}{\gamma} \frac{\partial^2 P}{\partial x^2}.
\end{equation}
After some algebra, the probability distribution reduces to
\begin{equation}
P(x,t) = \frac{1}{\sqrt{4\pi D t}} \exp \left( -\frac{(x - \frac{F}{\gamma} H(t - t_0))^2}{4 D t} \right)
\end{equation}
where $H(t)$ is the Heaviside step function. Force $F_0$ is applied at time $t_0$.

The entropy:
\begin{equation}
S(t) = \frac{1}{2} \ln \left( \frac{4\pi e T t}{\gamma L^2} \right)
\end{equation}
increases with time and temperature.

The velocity:
\begin{equation}
v = \frac{(F_0 - \gamma v_0) \sum_{n=0}^{\infty} \delta(t - t_0)}{\gamma}
\end{equation}
will have sudden jumps.

The entropy production rate $\dot{e}_p$ is calculated as
\begin{equation}
\dot{e}_p = \frac{1}{2 T t} + \frac{(F_0 - \gamma v_0)^2}{\gamma T} \delta(t - t_0).
\end{equation}
Similarly, the entropy extraction rate $\dot{h}_d$ is given by
\begin{equation}
\dot{h}_d = \frac{(F_0 - \gamma v_0)^2}{\gamma T} \delta(t - t_0).
\end{equation}
At steady state ($t \to \infty$), the term $\frac{1}{2 T t}$ vanishes, and we obtain
\begin{equation}
\dot{e}_p = \dot{h}_d = \frac{(F_0 - \gamma v_0)^2}{\gamma T} \delta(t - t_0)
\end{equation}
This confirms that the entropy production rate balances the entropy extraction rate.
As can be seen clearly, the entropy production rate is dictated by nonperiodic impulsive forces. When a sudden impulsive force is applied, abrupt spikes in the entropy production rate are observed. The spikes are then followed by relaxation phases at the same time that the rate decreases in time to a steady state. Despite the presence of transient fluctuations, the system relaxes back.

\begin{figure}[ht]
\centering
\includegraphics[width=6cm]{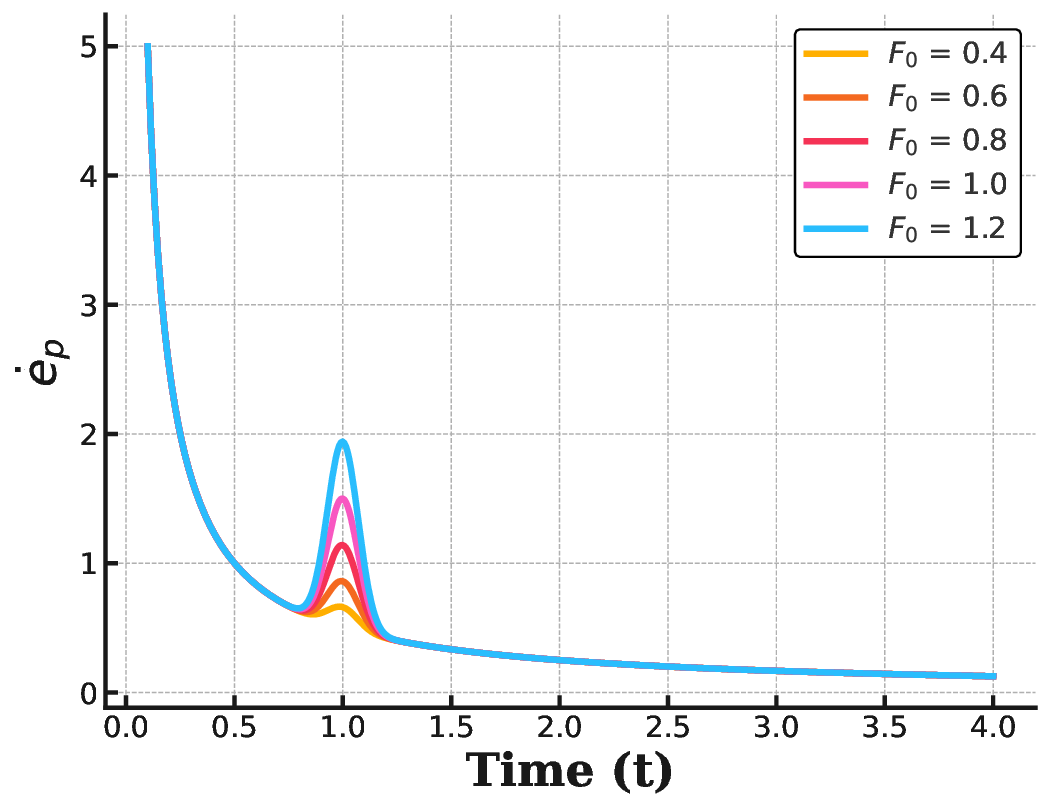}
\caption{The entropy production rate, $\dot{e}_p$ as function of time $t$ for fixed values of $F_0$ ($F_0 = 0.4, 0.6, 0.8, 1.0, 1.2$) and $v_0=0$ in the presence of an impulsive force. The results reveal that when a nonperiodic impulsive force is applied, the entropy production rate exhibits a sharp spike at a specific time. The system undergoes a relaxation period after this peak, during which the rate of entropy formation gradually decreases. As the magnitude of $F_0$ increases, the amplitude of the spike increases. In general, as time progresses, the entropy production or extraction rate decreases and saturates at a constant value.}
\label{fig:sub} 
\end{figure}

In Fig. 7, we plot the entropy production rate (when $v_0=0$), $\dot{e}_p$ as a function of time $t$ for fixed values of $F_0$ ($F_0 = 0.4, 0.6, 0.8, 1.0, 1.2$) in the presence of an impulsive force. The results reveal that when a nonperiodic impulsive force is applied, the entropy production rate exhibits a sharp spike at a specific time. After a pronounced peak, the system relaxes back. As the magnitude of $F_0$ increases, the amplitude of the spike increases. In general, as time progresses, the entropy production or extraction rate decreases and saturates to a constant value.

\section{ Time dependent    viscous friction }

Understanding time-dependent viscous friction along the reaction coordinate is essential for analyzing the thermodynamics of Brownian motors and molecular machines in nonequilibrium environments. Since viscosity modulates diffusion, motor velocity, and entropy production, its temporal variations significantly impact energy dissipation and transport efficiency. Increasing friction hinders motion, elevating dissipation and entropy production, whereas decreasing friction enhances efficiency and velocity. This study explores how time-dependent friction affects entropy and transport efficiency, offering insights into optimizing nanoscale motion and adaptive motor control.

Consider a Brownian particle that hops in a thermally uniform medium where the viscous friction is time dependent
\begin{eqnarray}
\gamma &=& \frac{1}{g(1 + t^z)}
\end{eqnarray}
and in this case the corresponding Fokker-Planck equation in overdamped medium is given as
\begin{eqnarray}
\frac{\partial P(x,t)}{\partial t} &=& \frac{\partial}{\partial x}\left(\frac{f - \gamma v_0}{\gamma(t)}\right)P(x,t)+\nonumber \\
&& \frac{\partial}{\partial x}\left(\frac{T}{\gamma(t)}\frac{\partial P(x,t)}{\partial x}\right).
\end{eqnarray}
Imposing a periodic boundary condition $P(0,t)=P(L,t)$ and let us choose a Fourier cosine series
\begin{eqnarray}
P(x,t) = \sum_{n=0}^\infty b_{n}(t) \cos\left(\frac{n\pi}{L}\left(x + \frac{f - \gamma v_0}{\gamma}\right)\right)
\end{eqnarray}
After some algebra we get the probability distribution as
\begin{eqnarray}
P(x,t) &=& \sum_{n=0}^\infty \cos\left[\frac{n\pi}{L}\left(x + (f - \gamma v_0)\left(gt + \frac{g t^{z+1}}{z+1}\right)\right)\right] \zeta
\end{eqnarray}
where
\begin{eqnarray}
\zeta &=& \exp\left(-\frac{(n\pi)^2 T \left(gt + \frac{g t^{z+1}}{z+1}\right)}{L^2}\right).
\end{eqnarray}
Here $f$ is the external load and $T$ is the temperature of the medium.

The current is then given by
\begin{eqnarray}
J(x,t) &=& -\left[\frac{(f - \gamma v_0) P(x,t)}{\gamma} + \frac{T}{\gamma} \frac{\partial P(x,t)}{\partial x}\right].
\end{eqnarray}

As stated before $\dot{e}_p = \dot{h}_d + \frac{d S(t)}{dt}$ where
\begin{eqnarray}
\frac{d S(t)}{dt} = -\int \frac{J}{P(x,t)} \frac{\partial P(x,t)}{\partial x} dx.
\end{eqnarray}
After some algebra, we write
\begin{widetext}
\begin{eqnarray}
\frac{d S(t)}{dt} &=& \int -J \frac{\sum{n=0}^\infty \frac{n\pi}{2} \cos\left[\frac{n\pi}{L}\left(x + (f - \gamma v_0)\left(gt + \frac{g t^{z+1}}{z+1}\right)\right)\right] \zeta}{\sum{n=0}^\infty \cos\left[\frac{n\pi}{L}\left(x + (f - \gamma v_0)\left(gt + \frac{g t^{z+1}}{z+1}\right)\right)\right] \zeta} dx.
\end{eqnarray}
\end{widetext}

For such a system we have
\begin{eqnarray}
\dot{e}_p &=& -\int \frac{J^2}{P(x,t) T g(1 + t^z)} dx
\end{eqnarray}
and after some algebra we find
\begin{equation}
\dot{e}_p = \frac{(f - \gamma v_0)^2 g (1 + t^z)}{T}.
\end{equation}
The entropy extraction rate is given as
\begin{eqnarray}
\dot{h}_d &=& -\int \left(\frac{J (f - \gamma v_0)}{T}\right) dx
\end{eqnarray}
and after some algebra one gets
\begin{equation}
\dot{h}_d = \frac{(f - \gamma v_0)^2 g (1 + t^z)}{T}.
\end{equation}

For $z<0$, in the limit $t \to \infty$, $\dot{e}_p = \dot{h}_d$. For the case where $z>0$, as $t$ increases, $\dot{e}_p$ and $\dot{h}_d$ monotonously increase.

On the other hand
\begin{eqnarray}
\Delta h_d = \Delta e_p &=& \int_0^t \frac{(f - \gamma v_0)^2 g (1 + t^z)}{T} dt \\ \nonumber
&=& \frac{(f - \gamma v_0)^2 g t (t^z + z + 1)}{T (z + 1)}.
\end{eqnarray}

The heat dissipation rate is given by
\begin{eqnarray}
\dot{H}_d &=& -\int (J (f - \gamma v_0)) dx \\ \nonumber
&=& (f - \gamma v_0)^2 g (1 + t^z).
\end{eqnarray}

The term $\dot{E}_p$ is related to $\dot{e}_p$ and it is given by
\begin{eqnarray}
\dot{E}_p &=& -\int \frac{J^2}{P(x,t) g(1 + t^z)} dx \\ \nonumber
&=& (f - \gamma v_0)^2 g (1 + t^z)
\end{eqnarray}
while
\begin{eqnarray}
\Delta H_d &=& \Delta E_p = \int_0^t (f - \gamma v_0)^2 g (1 + t^z) dt \\ \nonumber
&=& \frac{(f - \gamma v_0)^2 g t (t^z + z + 1)}{(z + 1)}.
\end{eqnarray}

On the other hand, the internal energy is given by
\begin{eqnarray}
\dot{E}_{in} = \int J U's(x) dx = 0.
\end{eqnarray}
The total work done is then given by
\begin{eqnarray}
\dot{W} = \int (J (f - \gamma v_0)) dx.
\end{eqnarray}
The first law of thermodynamics can be written as
\begin{eqnarray}
\dot{E}{in} = -\dot{H}_d(t) - \dot{W}.
\end{eqnarray}

\section{Optimal Driving Force to Minimize Entropy Production}

Optimizing the driving force to minimize entropy production is essential for improving energy efficiency and reducing unnecessary dissipation in nonequilibrium systems. In many physical and biological processes, such as molecular motors, nanoscale transport, and active matter, external forces drive the system away from equilibrium, leading to energy loss in the form of heat. If the applied force is not carefully controlled, excessive dissipation occurs, reducing the efficiency of energy conversion and making the system less stable over time.

By designing an optimal force protocol that dynamically adjusts to changes in temperature, the system can operate closer to equilibrium, reducing entropy production and minimizing energy waste. This optimization ensures that work is extracted efficiently while keeping dissipation as low as possible. It is particularly relevant in fields such as biophysics, where molecular machines rely on precise energy management, and in nanotechnology, where minimizing heat generation is crucial for device performance. Understanding and implementing optimal driving strategies help in designing energy-efficient processes, prolonging system stability, and improving the overall functionality of force-driven systems.

To minimize entropy production in a system with time-dependent temperature, an optimal driving force must be designed to reduce dissipation. Considering only the first term of the entropy production rate,  
\begin{equation}
\dot{e}_p = \frac{f^2}{T(t)},
\end{equation}
it is evident that entropy production is directly proportional to the square of the applied force and inversely proportional to the instantaneous temperature. To achieve minimal entropy generation, the driving force $f(t)$ must be adjusted dynamically in response to temperature variations.  

An optimal force protocol can be derived by enforcing a balance between energy input and dissipation constraints. One effective approach is to define $f(t)$ as a function of temperature such that the ratio $f^2 / T(t)$ remains as small as possible while maintaining system control. A natural choice is a power-law scaling,  
\begin{equation}
f_{\text{opt}}(t) = F_0 \left( \frac{T(t)}{T_0} \right)^\beta,
\end{equation}
where $F_0$ is the initial force magnitude, $T_0$ is the initial temperature, and $\beta$ is an adjustable parameter that determines how the force compensates for thermal fluctuations. By selecting an appropriate $\beta$, one can tailor the protocol to minimize entropy production while ensuring that the system remains in a controlled nonequilibrium state.  

Optimizing the driving force in this manner is crucial for improving energy efficiency in stochastic thermodynamic systems, particularly in molecular machines, nanoscale engines, and active matter. Proper force modulation reduces unnecessary dissipation, extends system longevity, and enhances performance in thermodynamic control processes,

\section{Summary and Conclusion}

In this work, we study the thermodynamics of Brownian motors  that operate in media where both the driving force and thermal background vary dynamically in time. While previous studies have mainly  focused on systems with isothermal or spatially varying forces or temperatures, real molecular machines such as intracellular transport proteins operate under temporally fluctuating forces and temperatures. These time-dependent protocols fundamentally alter the energy dissipation and entropy production, motivating a systematic theoretical investigation.

We analytically examined both active and passive Brownian motors subjected to exponentially, linearly, and quadratically varying thermal and force fields. The active motor is modeled as a particle self-propelling with velocity $v_0$, while the passive motor operates solely due to external driving and thermal asymmetries. Explicit expressions are derived for entropy, entropy production, and entropy extraction rates under each protocol.

Our results show that the total entropy depends solely on the temperature profile and viscous friction, provided that no boundary constraints are imposed. In contrast, entropy production and dissipation are directly modulated by the net driving force. In particular, we find that time-dependent periodic forces give rise to oscillatory entropy production superimposed on a monotonic decay, while periodic impulsive forces generate discrete entropy spikes. In contrast, nonperiodic impulses produce abrupt surges in entropy production, followed by relaxation toward a steady state. We also show that at stall force $f = \gamma v_0$,  thermodynamic rates such as entropy production and extraction rates  vanish. 

In conclusion, in this work, we  present an important model system that helps undrestand  the thermodynamic behavior of Brownian motors that  operate  under time-dependent forces and thermal arrangements. The analytical results derived herein  give   key insights into how nanoscale engines adapt to dynamic driving conditions.  This, in turn, give us   a theoretical foundation for the design and optimization of molecular machines capable of operating efficiently under fluctuating external stimuli.

\section*{Acknowledgment}
I would like to thank  Mulu  Zebene for the
constant encouragement. 

\section*{Data Availability Statement }This manuscript has no
associated data or the data will not be deposited. [Authors’
comment: Since we presented an analytical work, we did not
collect any data from simulations or experimental observations.]


\begin{thebibliography}{80}
\bibitem{mmg1} L. K. Davis, K. Proesmans, and É. Fodor, Phys. Rev. X {\bf 14}, 011012 (2024). 
\bibitem{mmg2} M. te Vrugt and R. Wittkowski, arXiv {\bf 2405.15751} (2024). 
\bibitem{mmg3} J. Mecke, J. O. Nketsiah, R. Li, and Y. Gao, Natl. Sci. Open {\bf 3}, 20230086 (2024). 
\bibitem{mmg4} L.-H. Cai, S. S. Datta, and X. Cheng, Front. Phys. {\bf 10}, 1005146 (2022). 
\bibitem{mu1} H. Ge and H. Qian, Phys. Rev. E {\bf 81}, 051133 (2010).
\bibitem{mu2} T. Tome and M. J. de Oliveira, Phys. Rev. Lett. {\bf 108}, 020601 (2012).
\bibitem{mu3} J. Schnakenberg, Rev. Mod. Phys. {\bf 48}, 571 (1976).
\bibitem{mu4} T. Tome and M.J. de Oliveira, Phys. Rev. E {\bf 82}, 021120 (2010).
\bibitem{mu5} R.K.P. Zia and B. Schmittmann, J. Stat. Mech. {\bf P07012} (2007). 
\bibitem{mu6} U. Seifert, Phys. Rev. Lett. {\bf 95}, 040602 (2005). 
\bibitem{mu7} T. Tome, Braz. J. Phys. {\bf 36}, 1285 (2006).
\bibitem{mu8} G. Szabo, T. Tome  and I. Borsos, Phys. Rev. E {\bf 82}, 011105 (2010).
\bibitem{mu9} B. Gaveau, M. Moreau and L.S. Schulman, Phys. Rev. E {\bf 79}, 010102 (2009).
\bibitem{mu10} J.L. Lebowitz and H. Spohn, J. Stat. Phys. {\bf 95}, 333 (1999).
\bibitem{mu11} D. Andrieux and P. Gaspar, J. Stat. Phys. {\bf 127}, 107 (2007). 
\bibitem{mu12} R.J. Harris and G.M. Schutz, J. Stat. Mech.  {\bf P07020} (2007). 
\bibitem{ta1}Tania Tome and Mario J. de Oliveira, Phys. Rev. E {\bf 9}, 042140 (2015).
\bibitem{mu13} J.-L. Luo, C. Van den Broeck, and G. Nicolis, Z. Phys. B {\bf 56}, 165 (1984). 
\bibitem{mu14} C.Y. Mou, J.-L. Luo, and G. Nicolis, J. Chem. Phys. {\bf 84}, 7011 (1986). 
\bibitem{mu15} C. Maes and K. Netocny, J. Stat. Phys. {\bf 110}, 269 (2003).
\bibitem{mu16} L. Crochik and T. Tome, Phys. Rev. E  {\bf 72}, 057103 (2005). 
\bibitem{mu17} M. Asfaw,  Phys. Rev. E {\bf 89}, 012143 (2014).
\bibitem{muu17} M. Asfaw,  Phys. Rev. E {\bf 92},  032126 (2015).
\bibitem{mu25} K. Brandner, M. Bauer, M. Schmid  and U. Seifert,  New. J. Phys. {\bf 17}, 065006 (2015).
\bibitem{mu26} B. Gaveau, M. Moreau and  L. S. Schulman, Phys. Rev. E {\bf 82}, 051109 (2010).
\bibitem{mu27} E. Boukobza and D.J. Tannor,  Phys. Rev. Lett. {\bf 98}, 240601 (2007).
\bibitem{mg40} M. A. Taye, Phys. Rev. E {\bf 110}, 054105 (2024).
\bibitem{mg41} M. A. Taye, Phys. Rev. E {\bf 105}, 054126 (2022).
\bibitem{mg22}  H. Ge, Phys. Rev. E {\bf 89}, 022127 (2014).
\bibitem{mg23} M. A. Taye and M. Bekele   Eur. Phys. J. B {\bf 38}, 457 (2004).
\bibitem{muuu17} M. A. Taye,  Phys. Rev. E {\bf 94},  032111 (2016).
\bibitem{muuu177} M. A. Taye,  Phys. Rev. E {\bf 101},  012131 (2020).  
\bibitem{mar2} H. Ge, Phys. Rev. E {\bf 89}, 022127 (2014).
\bibitem{mar1} H. K. Lee, C. Kwon, and H. Park, Phys. Rev. Lett. {\bf 110}, 050602 (2013).
\bibitem{mg6} R. E. Spinney and I. J. Ford, Phys. Rev. Lett. {\bf 108},170603 (2012).
\bibitem{mg7} H. K.Lee, C. Kwon and H. Park, Phys. Rev. Lett. {\bf 110}, 050602 (2013).
\bibitem{mg8} A. Celani, et al., Phys. Rev. Lett. {\bf 109}, 260603 (2012).
\bibitem{mg10} S. K. Manikandan, D. Gupta, and S. Krishnamurthy, Phys. Rev. Lett. {\bf 124}, 120603 (2020).
\bibitem{mg11} D. J. Skinner and J. DunkelPhys. Rev. Lett. {\bf 127}, 198101 (2021).
\bibitem{mg12} S. Otsubo, S. Ito, A. Dechant, and T. Sagawa, Phys. Rev. E {\bf 101}, 062106 (2020).
\bibitem{mg14} T. V. Vu, V. T. Vo, and Y. Hasegawa, Phys. Rev. E {\bf 101}, 042138 (2020).
\bibitem{mg15} T. Koyuk and U. Seifert, Phys. Rev. Lett. {\bf 122}, 230601 (2019).
\bibitem{mg9} P. Strasberg and M. Esposito, Phys. Rev. E {\bf 99}, 012120 (2019).
\end{thebibliography}
\end{document}